\UseRawInputEncoding
\documentclass[aps,prl,preprint,superscriptaddress]{revtex4-2}
\usepackage{graphicx}
\usepackage{dcolumn}
\usepackage{bm}
\usepackage{amsmath}
\usepackage{subfigure} 
\usepackage{amssymb}
\usepackage{latexsym}
\usepackage{epsfig}
\usepackage{amsbsy}
\usepackage{array}
\usepackage{comment}
\usepackage{booktabs}
\usepackage{xcolor}
\usepackage{ulem}
\begin{document}
\title{Quantum voting machine encoded with microwave photons}

\author{Yu Zhang}
\email[]{smutnauq@nju.edu.cn}
\affiliation{National Laboratory of Solid State Microstructures, School of Physics, Nanjing University, Nanjing 210093, China}
		\affiliation{Shishan Laboratory, Nanjing University, Suzhou 215163, China}
    \affiliation{Synergetic Innovation Center of Quantum Information and Quantum Physics, University of Science and Technology of China, Hefei, Anhui 230026, China}

\author{Chuiping Yang}
\affiliation{Department of Physics, Hangzhou Normal University, Hangzhou, Zhejiang 310036, China}

\author{Qiping Su}
\affiliation{Department of Physics, Hangzhou Normal University, Hangzhou, Zhejiang 310036, China}

\author{Yihao Kang}
\affiliation{Department of Physics, Hangzhou Normal University, Hangzhou, Zhejiang 310036, China}

\author{Wen Zheng}
\affiliation{National Laboratory of Solid State Microstructures, School of Physics, Nanjing University, Nanjing 210093, China}
		\affiliation{Shishan Laboratory, Nanjing University, Suzhou 215163, China}
    \affiliation{Synergetic Innovation Center of Quantum Information and Quantum Physics, University of Science and Technology of China, Hefei, Anhui 230026, China}

\author{Shaoxiong Li}
\affiliation{National Laboratory of Solid State Microstructures, School of Physics, Nanjing University, Nanjing 210093, China}
		\affiliation{Shishan Laboratory, Nanjing University, Suzhou 215163, China}
    \affiliation{Synergetic Innovation Center of Quantum Information and Quantum Physics, University of Science and Technology of China, Hefei, Anhui 230026, China}

\author{Yang Yu}
\email[]{yuyang@nju.edu.cn}
\affiliation{National Laboratory of Solid State Microstructures, School of Physics, Nanjing University, Nanjing 210093, China}
		\affiliation{Shishan Laboratory, Nanjing University, Suzhou 215163, China}
    \affiliation{Synergetic Innovation Center of Quantum Information and Quantum Physics, University of Science and Technology of China, Hefei, Anhui 230026, China}
    \affiliation{Hefei National Laboratory, Hefei 230088, China}


\date{\today}

\begin{abstract}
We propose a simple quantum voting machine using microwave photon qubit encoding, based on a setup comprising multiple microwave cavities and a coupled superconducting flux qutrit. This approach primarily relies on a multi-control single-target quantum phase gate. The scheme offers operational simplicity, requiring only a single step, while ensuring verifiability through the measurement of a single qubit phase information to obtain the voting results. And it provides voter anonymity, as the voting outcome is solely tied to the total number of affirmative votes. Our quantum voting machine also has scalability in terms of the number of voters. Additionally, the physical realization of the quantum voting machine is general and not limited to circuit QED. Quantum voting machine can be implemented as long as the multi-control single-phase quantum phase gate is realized in other physical systems. Numerical simulations indicate the feasibility of this quantum voting machine within the current quantum technology.
\end{abstract}

\maketitle

\section{}
The voting machine is a common decision-making tool widely used by various groups to vote on specific proposals. Generally, as long as it receives more than a certain proportion (usually 50\%) of the votes, the proposal can be passed and implemented. It is an indispensable tool in group decision-making. Anonymity and verifiability of the voting results are fundamental requirements for the voting machine. Recently, Some quantum voting schemes have been proposed\cite{li2021quantum,xu2022quantum,jiang2020quantum,xue2017simple,shi2021anonymous}, A novel quantum voting protocol using single-particle states is proposed to select multiple winners securely against malicious attacks\cite{li2021quantum}. A quantum voting protocol using non-orthogonal coherent states is proposed, eliminating the need for quantum memory storage\cite{xu2022quantum}, A quantum voting scheme is proposed based on locally indistinguishable orthogonal product states for efficient voting systems\cite{jiang2020quantum}. a simple quantum voting scheme\cite{xue2017simple} based on multi-particle entangled states is reported, it is suitable for large scale general votings and provides a proof of security against the most general type of attack. The proposed quantum voting protocol\cite{shi2021anonymous} ensures anonymity through the use of quantum operations and the chinese remainder theorem, making it infeasible for attackers to extract private information from partial qubits. However, a quantum voting machine (QVM) has not yet been implemented in circuit quantum electrodynamics (QED). Here, we propose a QVM based on circuit QED with microwave photon qubit encoding.\\ \indent
Circuit QED, formed through the integration of microwave cavities and superconducting (SC) qubits, has emerged as one of the leading candidates for quantum information processing \cite{yang2003possible, xiang2013hybrid,li2011quantum,wang2021improved}. After more than two decades of development, the coherence time of SC qubits has significantly improved from the initial nanosecond scale to hundreds of microseconds\cite{place2021new, wang2022towards} or even surpassing one millisecond\cite{yan2016flux, somoroff2023millisecond}. Furthermore, experiments have revealed the existence of one-dimensional microwave resonators characterized by a high quality factor ($Q\gtrsim10^{6}$) \cite{chen2008substrate, leek2010cavity, megrant2012planar, calusine2018analysis, woods2019determining, melville2020comparison}, as well as three-dimensional microwave cavities with a high quality factor ($Q\gtrsim3.5\times10^{7}$) \cite{reagor2016quantum, kudra2020high, romanenko2020three}. The experimental validation of these high quality factors suggests that microwave cavities or resonators contain microwave photons with lifetimes similar to those of SC qubits\cite{devoret2013superconducting}. The utilization of microwave fields and photons for quantum state engineering and quantum information processing has garnered considerable recent interest\cite{zhang2023generating, liu2022generation, zhang2017universal, puri2020bias, xu2022engineering, kang2022nonadiabatic, chen2022fault, krasnok2024superconducting, pietikainen2024strategies}.\\ \indent
\begin{figure}[htbp]
		\includegraphics[width=2.5in]{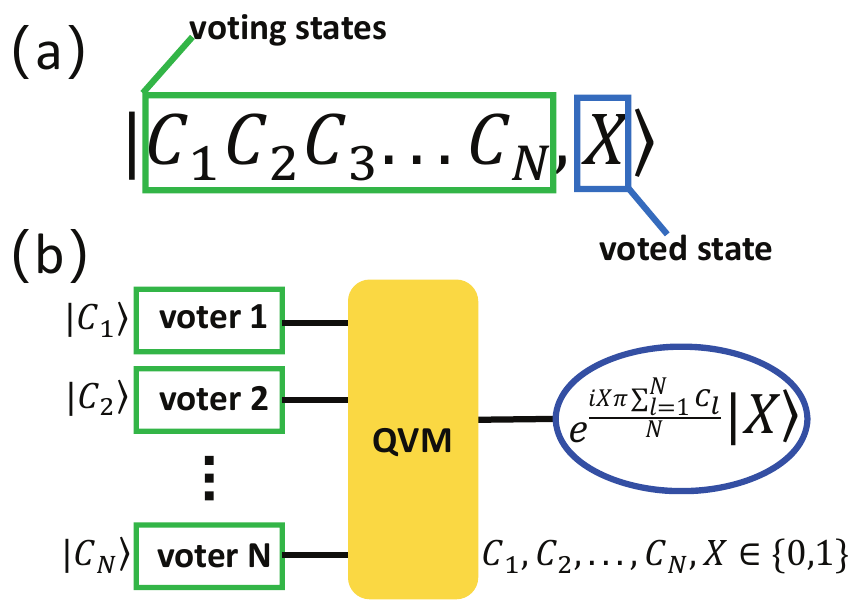}
		\caption{\label{fig:pic1-eps-converted-to} The schematic diagram illustrating the function of the QVM. (a) The composition of quantum state used in the QVM. (b) When $X = 1$, a phase related to the total number of affirmative votes is introduced into the quantum state.}

\end{figure}
\begin{figure*}[htbp]
		\includegraphics[width=4.50in]{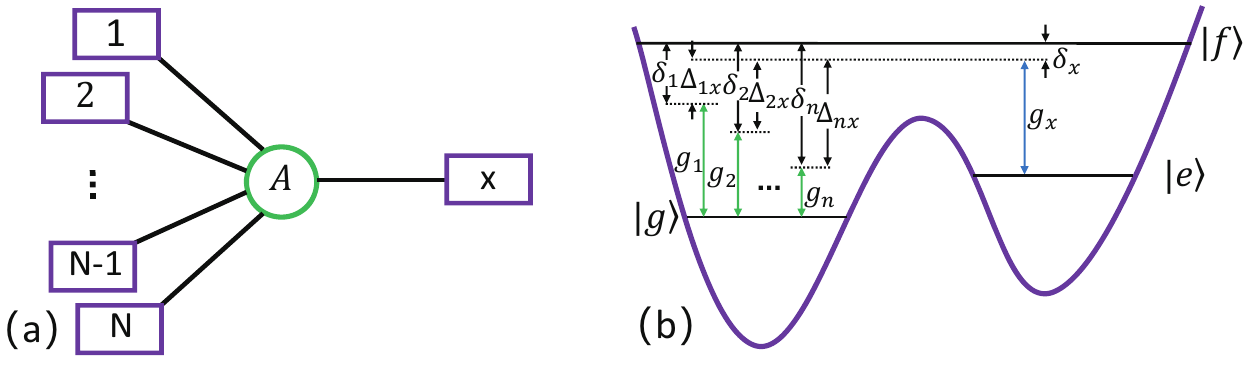}
		\caption{\label{fig:pic2-eps-converted-to} (a) Schematic circuit of $N+1$ microwave cavities coupled to a common flux qutrit. Each purple box represents a microwave cavity. The circle A represents the flux qutrit, which is inductively or capacitively coupled to each cavity. (b) Level configuration of the flux qutrit, for which the transition between the two lowest levels can be made weak by increasing the barrier between two potential wells.}
\end{figure*}
As depicted in Fig. \ref{fig:pic1-eps-converted-to}, the function of the QVM is to determine the total number of voters in favor of the event being voted upon at the conclusion of the voting process. In Fig. \ref{fig:pic1-eps-converted-to} (a), the voters can be denoted as $|C_1 C_2 C_3...C_N\rangle$, where $C_1,...,C_N \in \{0, 1\}$; and the state $|1\rangle$ represents affirmative vote, while the state $|0\rangle$ represents negative vote. The event being voted upon can be denoted as $|X\rangle$. As shown in Fig. \ref{fig:pic1-eps-converted-to} (b), when $X$ is $0$, the quantum state of the system remains unchanged. When $X$ is $1$, the system's quantum state acquires a phase related to the state of each voter. The function of QVM is given by the following equation:
\begin{equation}\label{eq:func0}
\begin{aligned}
& U\left|C_1, C_2, \ldots, C_N, 0\right\rangle=\left|C_1, C_2, \ldots, C_N, 0\right\rangle, \\
& U\left|C_1, C_2, \ldots, C_N, 1\right\rangle=e^{i \frac{\pi\sum^{N}_{l=1}{C_l}}{N}}\left|C_1, C_2, \ldots, C_N, 1\right\rangle,
\end{aligned}
\end{equation}
where $U$ is actually a multi-control single-target quantum phase gate. And we implement this QVM based on such a device, multiple microwave cavities coupled to a common flux qutrit [Fig. \ref{fig:pic2-eps-converted-to} (a)]. The $N$ microwave cavities on the left represent $N$ voters, while the one on the right represents the event to be voted on. The three levels of the qutrit are labeled as $|g\rangle$, $|e\rangle$, and $|f\rangle$ [Fig. \ref{fig:pic2-eps-converted-to}(b)]. Suppose that cavity $l$ $( l = 1, 2, ..., N )$ are dispersively coupled to the $|g\rangle\leftrightarrow|f\rangle$ transition with coupling constant $g_{l}$ and detuning $\delta_{l}$ but highly detuned (decoupled) from the $|e\rangle\leftrightarrow|f\rangle$ transition of the qutrit. In addition, assume that cavity $x$ is dispersively coupled to the $|e\rangle\leftrightarrow|f\rangle$ transition with coupling constant $g_{x}$ and detuning $\delta_{x}$ but highly detuned (decoupled) from the $|g\rangle\leftrightarrow|f\rangle$ transition of the qutrit [Fig. \ref{fig:pic2-eps-converted-to}(b)]. These conditions can be met by prior adjustment of the level spacings of the qutrit or the frequencies of the cavities. Note that both of the level spacings of a SC qutrit and the frequency of a microwave cavity can be rapidly tuned within a few nanoseconds\cite{neeley2008process,sun2010tunable,wang2013quantum}.\\ \indent
After the above considerations, the Hamiltonian of the whole system in the interaction picture and under the rotating wave approximation is given by (assuming $\hbar=1$)
\begin{equation}\label{eq:H_I}
H_I=\sum_{l=1}^N\left(g_l e^{-i \delta_l t} \hat{a}_l^{+} \delta_{f g}^{-}+\mathrm{H.c.}\right)+g_x e^{-i \delta_x t} \hat{a}_x^{+} \delta_{f e}^{-}+\mathrm{H.c.},
\end{equation}
where $\hat{a}_{l}$ $(\hat{a}_{x})$ is the photon annihilation operator of cavity $l$ (cavity $x$), $\sigma_{f g}^{-}=|g\rangle\langle f|$, $\sigma_{f e}^{-}=|e\rangle\langle f|$, $\delta_l=\omega_{f g}-\omega_{c_l}>0$, and $\delta_x=\omega_{f e}-\omega_{c_x}>0$ [Fig. \ref{fig:pic2-eps-converted-to}(b)]. Here, $\omega_{fg}$ ($\omega_{fe}$) is the $|f\rangle\leftrightarrow|g\rangle$ ($|f\rangle\leftrightarrow|e\rangle$) transition frequency of the qutrit, while $\omega_{c_l}$ ($\omega_{c_x}$) is the frequency of cavity $l$ ($x$).\\ \indent
Under the large detuning conditions $\delta_l \gg g_l$ $(l = 1,2,3,...,N)$ and $\delta_x \gg g_x$, the energy exchange between the coupler qutrit and the cavities is negligible. In addition, under the condition of
\begin{equation}
\frac{\left|\delta_l-\delta_{l^{'}}\right|}{\delta_l^{-1}+\delta_{l^{'}}^{-1}} \gg g_l g_{l^{'}},
\end{equation}
(where $l$, $l^{'}$ $\in\{1, 2,..., N\}$ and $l\neq l^{'}$), the interaction between the cavities (1, 2, ..., N), induced by the coupler qutrit, is negligible. Therefore, the Hamiltonian (\ref{eq:H_I}) becomes\cite{zheng2000efficient,sorensen1999quantum,james2007effective}
\begin{equation}\label{eq:HE0}
\begin{aligned}
H_e= & \sum_{l=1}^N\left(-\lambda_l \widehat{n}_l|g\rangle\left\langle g\left|+\lambda_l\left(1+\widehat{n}_l\right)\right| f\right\rangle\langle f|\right) \\
& -\lambda_x \widehat{n}_x|e\rangle\left\langle e\left|+\lambda_x\left(1+\widehat{n}_x\right)\right| f\right\rangle\langle f| \\
& -\sum_{l=1}^N \lambda_{l x}\left(e^{i \Delta_{lx} t} \hat{a}_l \hat{a}_x^{+} \sigma_{g e}^{-}+\mathrm{H.c.}\right),
\end{aligned}
\end{equation}
where $\sigma_{e g}^{-}=|g\rangle\langle e|$, $\lambda_l=g_l^2 / \delta_l$, $\lambda_x=g_x^2 / \delta_x$, $\lambda_{l x}=$ $\left(g_l g_x / 2\right)\left(1 / \delta_l+1 / \delta_x\right)$, $\Delta_{l x}=\delta_l-\delta_x=\omega_{eg}+\omega_{c_x}-\omega_{c_l}$. In Eq. (\ref{eq:HE0}), the terms in the first two lines describe the photon-number-dependent stark shifts of the energy levels $|g\rangle$, $|e\rangle$, and $|f\rangle$, while the terms in the last line describe the $|g\rangle\leftrightarrow|e\rangle$ coupling caused due to the cooperation of cavities $l$ and $x$. When $\Delta_{l x} \gg\left\{\lambda_l, \lambda_x, \lambda_{l x}\right\}$, the effective
Hamiltonian $H_e$ turns into \cite{zheng2000efficient,sorensen1999quantum,james2007effective}
\begin{equation}\label{eq:HE1}
\begin{aligned}
H_e= & \sum_{l=1}^N(-\lambda_l \widehat{n}_l|g\rangle\langle g|+\lambda_l(1+\widehat{n}_l)| f\rangle\langle f|)-\lambda_x \widehat{n}_x|e\rangle\langle e|\\
&+\lambda_x(1+\widehat{n}_x)| f\rangle\langle f|-\sum_{l=1}^n \chi_{l x} \widehat{n}_l(1+\widehat{n}_x)|g\rangle\langle g| \\
& +\sum_{l=1}^N \chi_{l x}(1+\widehat{n}_l) \widehat{n}_x|e\rangle\langle e|,
\end{aligned}
\end{equation}
where $\chi_{l x}=\lambda_{l x}^2 / \Delta_{l x}$. In the case of the levels $|e\rangle$ and $|f\rangle$ being initially not occupied, these levels will not be populated because neither $|g\rangle\leftrightarrow|e\rangle$ transition nor $|g\rangle\leftrightarrow|f\rangle$ transition is induced by the Hamiltonian (\ref{eq:HE1}). Therefore, the Hamiltonian (\ref{eq:HE1}) reduces to
\begin{equation}\label{eq:HE2}
H_e=  -\sum_{l=1}^N \lambda_l \widehat{n}_l|g\rangle\langle g|-\sum_{l=1}^N \chi_{l x} \widehat{n}_l\left(1+\widehat{n}_x\right)|g\rangle\langle g|,
\end{equation}
where . In the following, we assume that the qutrit is initially in the ground state $|g\rangle$. Note that the qutrit will remain in this state because none of interlevel transitions of the qutrit can be induced by the Hamiltonian (\ref{eq:HE2}). The Hamiltonian (\ref{eq:HE2}) thus reduces to
\begin{equation}\label{eq:HE3}
\widetilde{H}_e= -\sum_{l=1}^N \eta_{l} \widehat{n}_l-\sum_{l=1}^N \chi_{l x} \widehat{n}_l \widehat{n}_x,
\end{equation}
where $\eta_l = \lambda_l +\chi_{l x}$. The effective Hamiltonian $\widetilde{H}_{e}$ characterizes the dynamics of the cavity system. Under this Hamiltonian, the unitary operator $U=e^{-i \widetilde{H}_{\mathrm{e}} t}$ describing the state time evolution of the cavity system can be expressed as
\begin{equation}
U=\prod_{l=1}^N U_{l x} \otimes \prod_{l=1}^N U_l
\end{equation}
with
\begin{equation}
U_{l x}=\exp \left(i \chi_{l x} \widehat{n}_l \widehat{n}_x t\right),\quad U_l=\exp \left(i \eta_l \widehat{n}_l t\right),
\end{equation}
where $U_{l}$ is a unitary operator on cavity $l$, while $U_{lx}$ is a unitary operator on cavities $x$ and $l$ $(l = 1, 2, ..., N)$. For two conditions below
\begin{equation}\label{eq:s_l}
\eta_{l}t = 2\pi s_{l},\quad\chi_{lx} t=\frac{\pi}{N},
\end{equation}
where $s_{l}$ is a positive integer for cavity $l$, $N$ is the total number of control qubits. The unitary operator $U$ of the cavity system can be finally expressed as
\begin{equation}\label{eq:U}
U = \exp \left(i \frac{\pi}{N} \widehat{n}_x \sum_{l=1}^N \widehat{n}_l\right).
\end{equation}
For $N+1$ photonic qubits, target qubit $x$ and control qubit $l$ $(l = 1, 2, ..., N)$, we consider the initial state of the cavity system $|n_1, n_2, ..., n_N, n_x\rangle_{m}$ $(n_1, n_2, n_3, ..., n_N, n_x \in \{0, 1\})$, where subscript $m$ is the number of the control qubits in the state $|1\rangle$.
After the unitary operator $U$ (given in Eq. \ref{eq:U}), it is straightforward to get the following state transformation:
\begin{equation}
\begin{aligned}
&U |n_1, n_2, ..., n_N, 0\rangle_{m} = |n_1, n_2, ..., n_N, 0\rangle_{m},\\
&U |n_1, n_2, ..., n_N, 1\rangle_{m} = e^{i\frac{m\pi}{N}}|n_1, n_2, ..., n_N, 1\rangle_{m},
\end{aligned}
\end{equation}
which implies that if the target qubit $x$ is in the state $|0\rangle$, the state of the cavity system remains unchanged; while, if the target qubit $x$ is in the state $|1\rangle$, the phase associated with $m$ appears in the final state. When the target qubit $x$ is in the quantum state $|+\rangle=(|0\rangle+|1\rangle)/\sqrt{2}$, after the vote, a relative phase related to $m$ is introduced between the state $|0\rangle$ and the state $|1\rangle$ in the quantum state of the target qubit $x$. This phase can be obtained by performing Wigner tomography\cite{shalibo2013direct} on the target qubit.\\ \indent
\begin{figure}[htbp]
		\includegraphics[width=2.5in]{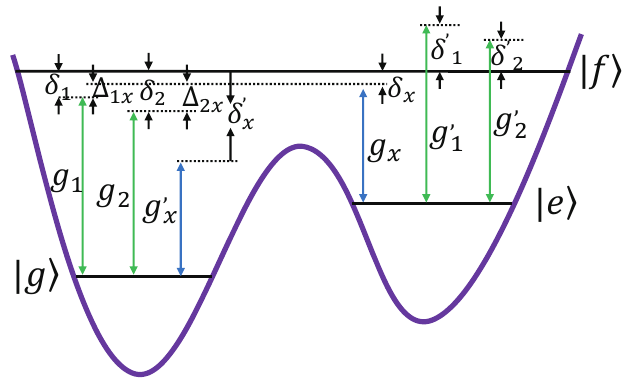}
		\caption{\label{fig:pic3-eps-converted-to} The representation of the coupling between cavity $1$ and the $|g\rangle \leftrightarrow |f\rangle$ transition of the qutrit (with coupling constant $g_1$ and detuning $\delta_1$), the coupling between cavity $2$ and the $|g\rangle \leftrightarrow |f\rangle$ transition of the qutrit (with coupling constant $g_{2}$ and detuning $\delta_{2}$), and the coupling between cavity $x$ and the $|e\rangle \leftrightarrow |f\rangle$ transition of the qutrit ( with coupling constant $g_x$ and detuning $\delta_{x}$). Furthermore, depiction of the unwanted coupling between cavity $1$ and the $|e\rangle \leftrightarrow |f\rangle$ transition of the qutrit (with coupling constant $g^{'}_{1}$ and detuning $\delta^{'}_{1}$ ), the unwanted coupling between cavity $2$ and the $|e\rangle \leftrightarrow |f\rangle$ transition of the qutrit (with coupling constant $g^{'}_{2}$ and detuning $\delta^{'}_{2}$), as well as the unwanted coupling between cavity $x$ and the $|g\rangle \leftrightarrow |f\rangle$ transition of the qutrit (with coupling constant $g^{'}_{x}$ and detuning $\delta^{'}_{x}$). Note that the coupling of each cavity with the $|g\rangle \leftrightarrow |e\rangle$ transition of the qutrit is negligible due to the weak $|g\rangle \leftrightarrow |e\rangle$ transition.}
\end{figure}
As an example, we investigate the experimental feasibility for the simple 2-voter QVM by using three one-dimensional microwave cavities coupled to a SC flux qutrit. the initial state of the whole system is
\begin{equation}
|\psi(0)\rangle=|n_1, n_2, +\rangle\otimes|g\rangle \quad(n_1, n_2\in\{0, 1\}),
\end{equation}
where $|g\rangle$ is the ground state of the qutrit $A$, and $|+\rangle=(|0\rangle+|1\rangle)/\sqrt{2}$ is the state of cavity $x$, which is the equal probability superposition of the vacuum state $|0\rangle$ and the single-photon sate $|1\rangle$. After applying the unitary operator $U$ of Eq. $\ref{eq:U}$ with two control qubits $(N=2)$, the initial state changes to
\begin{equation}
|\psi(T)\rangle=|n_1, n_2\rangle \otimes \frac{1}{\sqrt{2}}(|0\rangle+e^{i\frac{m\pi}{2}}|1\rangle)\otimes |g\rangle \quad(n_1, n_2\in\{0, 1\}),
\end{equation}
where only a phase $m\pi/2$ $(m = n_1 +n_2)$ associated with two control qubits is attached to the state $|1\rangle$ of the target qubits, while the state $|0\rangle$ of the target qubit remains unchanged.\\ \indent
The Hamiltonian used for the generation of $|\psi(T)\rangle$ is the effective Hamiltonian (\ref{eq:HE3}) with $N = 2$. This Hamiltonian was derived from the original Hamiltonian (\ref{eq:H_I}) with $N = 2$, which only includes the coupling between cavity $l$ $(l = 1, 2)$ and the $|g\rangle\leftrightarrow|f\rangle$ transition as well as the coupling between cavity $x$ and the $|e\rangle\leftrightarrow|f\rangle$ transition of the SC flux qutrit. In a realistic situation as showed in Fig. (\ref{fig:pic3-eps-converted-to}), there are the unwanted coupling between cavity $l$ $(l=1, 2)$ and the $|e\rangle\leftrightarrow|f\rangle$ transition as well as the unwanted coupling between cavity $x$ and the $|g\rangle\leftrightarrow|f\rangle$ transition of the SC flux qutrit. Besides, there is the unwanted intercavity crosstalk between the three cavities. When the unwanted couplings and the unwanted intercavity crosstalk are taken into account, the Hamiltonian (\ref{eq:H_I}), with $N = 2$ for the present case, is modified as
\begin{widetext}
\begin{equation}\label{eq:HI0}
\begin{aligned}
H_{I_{0}}=&\sum_{l=1}^{2}\left(g_{l} e^{-i \delta_{l} t} \hat{a}_{l}^{+} \delta_{f g}^{-}+\mathrm{H.c.}\right)+g_{x} e^{-i \delta_{x} t} \hat{a}_{x}^{+} \delta_{f e}^{-}+\mathrm{H.c.} \\&+\sum_{l=1}^{2}\left(g_{l}^{\prime} e^{-i \delta_{l}^{\prime} t} \hat{a}_{l}^{+} \delta_{f e}^{-}+\mathrm{H.c.}\right)+g_{x}^{\prime} e^{-i \delta_{x}^{\prime} t} \hat{a}_{x}^{+} \delta_{f g}^{-}+\mathrm{H.c.} \\&+\left(\widetilde{g}_{12} e^{-i \widetilde{\Delta}_{12} t} \hat{a}_{1}^{+} \hat{a}_{2}+\widetilde{g}_{2 x} e^{-i \widetilde{\Delta}_{2 x} t} \hat{a}_{2}^{+} \hat{a}_{x}+\widetilde{g}_{1 x} e^{-i \widetilde{\Delta}_{1 x} t} \hat{a}_{1}^{+} \hat{a}_{x}+\mathrm{H.c.}\right),
\end{aligned}
\end{equation}
\end{widetext}
where the terms in line one is the ideal Hamiltonian (\ref{eq:H_I}) with $N = 2$, the first term in line two represents the unwanted coupling between cavity $l$ $(l = 1, 2)$ and the $|e\rangle\leftrightarrow|f\rangle$ transition of the SC qutrit with coupling
constant $g^{'}_{l}$ and detuning $\delta^{'}_{l}=\omega_{fe}-\omega_{c_l}$, the second term in line two represents the unwanted coupling between cavity $x$ and the $|g\rangle\leftrightarrow|f\rangle$ transition of the SC qutrit with coupling constant $g^{'}_{x}$ and detuning $\delta^{'}_{l}=\omega_{fg}-\omega_{c_x}$, while the terms in the last line represent the unwanted intercavity crosstalk among the three
cavities, with $\tilde{g}_{ll^{'}}$ $(\tilde{\triangle}_{ll^{'}}=\omega_{l}-\omega_{l^{'}})$ the crosstalk strength (the frequency detuning) between the two cavities $l$ and $l^{'}$ $(l, l^{'}=1, 2, x; l\neq l^{'})$. Because the $|g\rangle\leftrightarrow|e\rangle$ transition can be made weak by increasing the barrier between the two potential wells of the SC qutrit, the coupling of each cavity with the $|g\rangle\leftrightarrow|e\rangle$ transition of the SC qutrit is negligible and thus is not considered in the above Hamiltonian (\ref{eq:HI0}).\\ \indent
Taking into account finite qutrit relaxation, dephasing, and photon lifetime, the dynamics of the lossy system is governed by the following master equation:
\begin{equation}\label{eq:master}
\begin{aligned}
\frac{d \rho}{d t}= & -i\left[H_{I_0}, \rho\right]+\sum_{j=1}^3 \kappa_j \mathcal{L}\left[\hat{a}_j\right] \\
& +\gamma_{e g} \mathcal{L}\left[\sigma_{e g}^{-}\right]+\gamma_{f e} \mathcal{L}\left[\sigma_{f e}^{-}\right]+\gamma_{f g} \mathcal{L}\left[\sigma_{f g}^{-}\right] \\
& +\gamma_{e, \varphi}\left(\sigma_{e e} \rho \sigma_{e e}-\sigma_{e e} \rho / 2-\rho \sigma_{e e} / 2\right) \\
& +\gamma_{f, \varphi}\left(\sigma_{f f} \rho \sigma_{f f}-\sigma_{f f} \rho / 2-\rho \sigma_{f f} / 2\right),
\end{aligned}
\end{equation}
where $\mathcal{L}[\Lambda]=\Lambda \rho \Lambda^{+}-\Lambda^{+} \Lambda \rho / 2-\rho \Lambda^{+} \Lambda / 2$ $(\Lambda=\hat{a}_j, \sigma_{e g}^{-}, \sigma_{f e}^{-}, \sigma_{f g}^{-})$, $\sigma_{e e}=|e\rangle\langle e|$, $\sigma_{ff}=|f\rangle\langle f|$, $\gamma_{e g}$ is the energy relaxation rate of the level $|e\rangle$ for the decay path $|e\rangle \rightarrow|g\rangle, \gamma_{f e}\left(\gamma_{f g}\right)$ is the relaxation rate of the level $|f\rangle$ for the decay path $|f\rangle \rightarrow|e\rangle$ $(|f\rangle \rightarrow|g\rangle)$, $\gamma_{e, \varphi}(\gamma_{f, \varphi})$  is the dephasing rate of the level $|e\rangle(|f\rangle)$ of the qutrit, while $\kappa_{j}$ is the decay rate of cavity $j$ $(j=1,2,x)$. We utilize the QUTIP\cite{johansson2012qutip} for numerical calculations, which is an open-source tool designed for simulating the dynamics of open quantum systems. \\ \indent
To measure the accuracy of the QVM, we evaluate the fidelity of the final state $|\psi(T_f)\rangle$ by
\begin{equation}
\mathcal{F}=\sqrt{\left\langle\psi_{\mathrm{id}}|\rho| \psi_{\mathrm{id}}\right\rangle},
\end{equation}
where $|\psi_{\mathrm{id}}\rangle=|\psi(T_f)\rangle$ the ideal final state, The ideal final state $|\psi_{\mathrm{id}}\rangle$ is derived without considering in system dissipation, intercavity crosstalk, and unwanted couplings, whereas $\rho$ represents the density operator characterizing the realistic final state of the system, which is achieved by numerically solving the master equation and considering the operation performed in a realistic situation.\\ \indent
\begin{table}\scriptsize
\caption{\label{tab:table1}Parameters used in the numerical simulation.}
\begin{ruledtabular}
\begin{tabular}{lcr}
$\omega_{g e} / 2 \pi$ = $8.0 \mathrm{GHz}$ & $\omega_{e f} / 2 \pi$ = $12.0 \mathrm{GHz}$ & $\omega_{g f} / 2 \pi$ = $20.0 \mathrm{GHz}$\\
\\
$\omega_{c 1} / 2 \pi$ = $19.159 \mathrm{GHz}$ & $\omega_{c 2} / 2 \pi$ = $19.120 \mathrm{GHz}$ & $\omega_{c x} / 2 \pi=11.208 \mathrm{GHz}$ \\
\\
$\delta_1 / 2 \pi$ = $841 \mathrm{MHz}$ & $\delta_2 / 2 \pi$ = $880 \mathrm{MHz}$ & $\delta_x / 2 \pi$ = $792 \mathrm{MHz}$ \\
\\
$\delta_1^{\prime} / 2 \pi$ = $-7.159 \mathrm{GHz}$ & $\delta_2^{\prime} / 2 \pi$ = $-7.120 \mathrm{GHz}$ & $\delta_x^{\prime} / 2 \pi$ = $8.792 \mathrm{GHz}$ \\
\\
$\widetilde{\Delta}_{12} / 2 \pi$ = $39 \mathrm{MHz}$ & $\widetilde{\Delta}_{1 x} / 2 \pi$ = $7.951 \mathrm{GHz}$ & $\widetilde{\Delta}_{2 x} / 2 \pi$ = $7.912 \mathrm{GHz}$ \\
\\
$g_1 / 2 \pi=21.4 \mathrm{MHz}$ & $g_2 / 2 \pi=29.4 \mathrm{MHz}$ & $g_x / 2 \pi=88 \mathrm{MHz}$ \\
\\
$g_1^{\prime} / 2 \pi$ = $21.4 \mathrm{MHz}$ & $g_2^{\prime} / 2 \pi$ = $29.4 \mathrm{MHz}$ & $g_x^{\prime} / 2 \pi$ = $88 \mathrm{MHz}$\\
\\
$g_{12}/2\pi= 0.88\mathrm{MHz}$ & $g_{1x}/2\pi= 0.88\mathrm{MHz}$ & $g_{2x}/2\pi= 0.88\mathrm{MHz}$\\
\\
$\kappa_{1} = \kappa_{2} = \kappa_{x} = \kappa$ & $\gamma^{-1}_{fe} = T$ & $\gamma^{-1}_{fg} = T$\\
\\
$\gamma^{-1}_{eg} = 10T$ & $\gamma^{-1}_{\phi e} = T/2$ & $\gamma^{-1}_{\phi f} = T/2$
\end{tabular}
\end{ruledtabular}
\end{table}
\begin{figure*}[htbp]
		\includegraphics[width=5.0in]{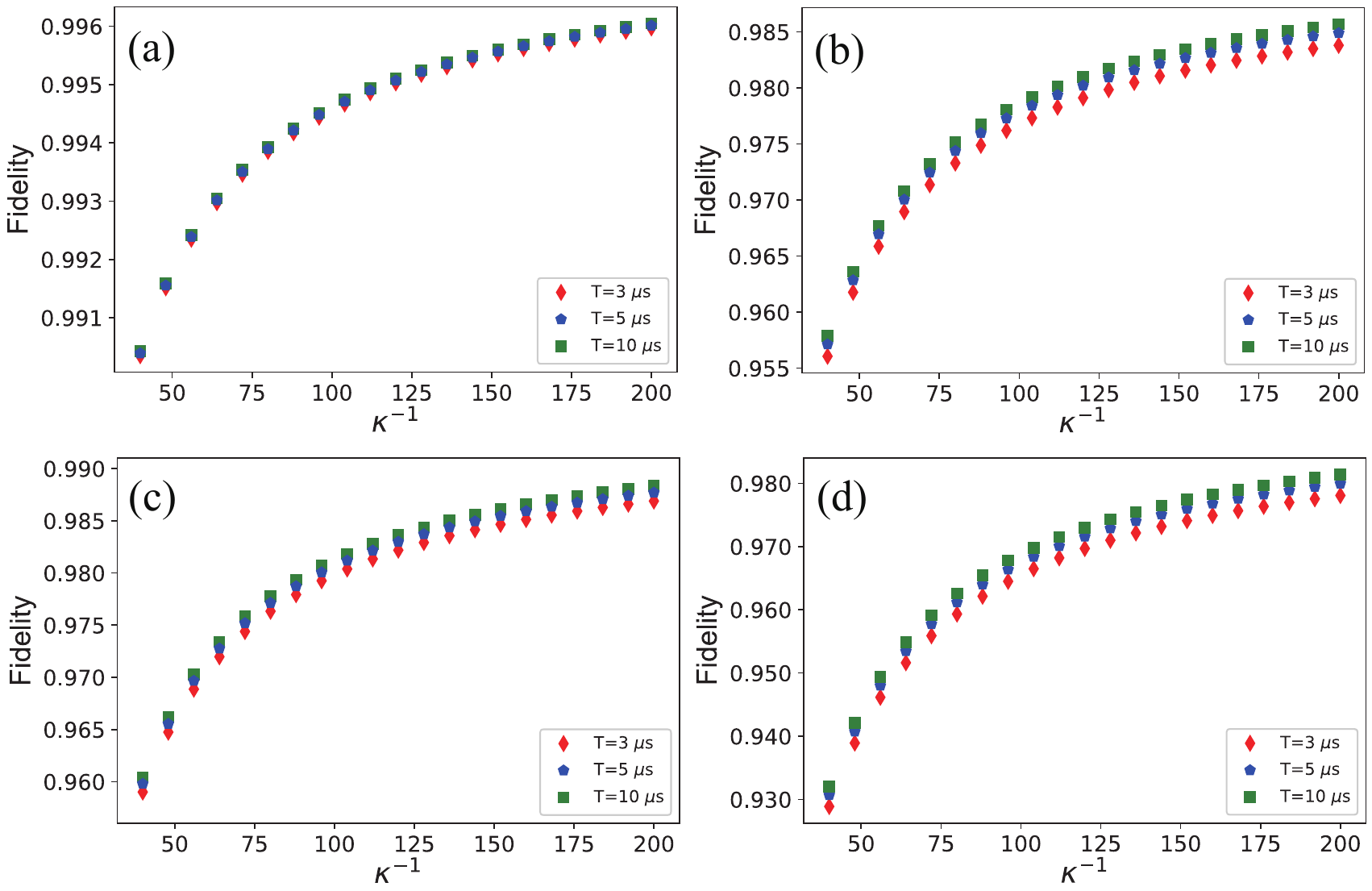}
		\caption{\label{fig:pic4-eps-converted-to}  Fidelity versus $\kappa^{-1}$ for $T=3$, $5$, and $10$ $\mu$s. (a) the situation where two voters cast negative votes with an initial state $|00,+\rangle$. (b) the situation where two voters each cast only one affirmative vote with an initial state $|01,+\rangle$. (c) the situation where two voters each cast only one affirmative vote with an intial state $|10,+\rangle$. (d) the situation where both voters cast affirmative votes with an initial state $|11,+\rangle$.}
	\end{figure*}
Table I lists the parameters used in the numerical simulations. By a proper design of the flux qutrit \cite{liu2005optical}, one can have $\phi_{f g} \sim \phi_{f e} \sim 10 \phi_{e g}$, where $\phi_{i j}$ is the dipole coupling matrix element between the two levels $|i\rangle$ and $|j\rangle$ with $ij\in\{e g, f e, f g\}$. Thus, one has $g_1^{\prime} \sim g_1$, $g_2^{\prime} \sim g_2$ and $g_x^{\prime} \sim g_x$, while the $|g\rangle\leftrightarrow|e\rangle$ transition is much weaker compared to the $|g\rangle\leftrightarrow|f\rangle$ and $|e\rangle\leftrightarrow|f\rangle$ transitions of the qutrit. Other parameters used in the numerical simulations are as follows: (i) $\gamma_{e g}^{-1}=10 T$, $\gamma_{f e}^{-1}=T$, $\gamma_{f g}^{-1}=T$, $\gamma_{\phi e}^{-1}=\gamma_{\phi f}^{-1}= T/2$, (ii) $\kappa_1=\kappa_2=\kappa_3=\kappa$, (iii) $g_{12}=g_{2x}=g_{1x}=g_{c r}$. We should mention that $\gamma_{e g}^{-1}$ is much larger than $\gamma_{f e}^{-1}$ and $\gamma_{f g}^{-1}$ because of $\phi_{f g} \sim \phi_{f e} \sim 10 \phi_{e g}$. The value of $T$ adopted in our numerical simulations is 3 $\sim$ 10 $\mu$s. Hence, the decoherence time of the qutrit used in the numerical simulations ranges from 30 to 100 $\mu$s. It is a rather conservative consideration, as experimental evidence has demonstrated decoherence times of 70 $\mu$s to 1 $m$s for SC flux devices\cite{yan2016flux,you2007low}.\\ \indent
By numerically solving the master Equation (\ref{eq:master}), we plot Fig. \ref{fig:pic4-eps-converted-to} to present the fidelity versus $\kappa^{-1}$ for $T=3$ $\mu$s (red rhombus), $5$ $\mu$s (blue pentagon), $10$ $\mu$s (green square), $g_{c r}=0.01$ max\{$g_1$, $g_2$, $g_x$\} and $s_1 (s_2)=1,(2)$ for Eq. (\ref{eq:s_l}). Fig. \ref{fig:pic4-eps-converted-to} (a) shows the situation where two voters cast negative votes with an initial state $|00,+\rangle$. Obviously, for this initial state, the photon number in the two cavities of the control qubits is $0$. The fidelity is thus less affected by the cavity dissipation compared to (b), (c), and (d). Furthermore, Fig. \ref{fig:pic4-eps-converted-to} (a) shows that the fidelity is also only slightly influenced by changes in $T$. Fig. \ref{fig:pic4-eps-converted-to} (b) and (c) respectively show the situation where two voters each cast only one affirmative vote with intial state $|01,+\rangle$ and $|10,+\rangle$. In these two cases, the photon number only in one of the two cavities (control qubits), and the fidelity is significantly more affected by the cavity dissipation compared to (a). Additionally, Fig. \ref{fig:pic4-eps-converted-to} (b) and \ref{fig:pic4-eps-converted-to} (c) demonstrate that the fidelity increases as $T$ increases. Fig. \ref{fig:pic4-eps-converted-to} (d) shows the situation where both voters cast affirmative votes with an initial state $|11,+\rangle$. The photon number in either of the two cavities (control qubits) is 1. The effect of the cavity dissipation is significantly greater than that in other cases. For example, at $T=3$ $\mu$s and $\kappa^{-1}=40$ $\mu$s, the fidelity in case (d) is lower than 93\%, while the fidelity in cases (b) and (c) is higher than 95.5\%, and it is even higher than 99\% in case (a).\\ \indent
In summary, We have proposed a scheme for implementing a QVM based on the microwave photonic qubit encoding. In this scheme, the QVM primarily operates the voting process through a multi-control single-target quantum phase gate. Due to the fact that the voting result is only related to the total number of affirmative votes, and thus incapable of reflecting specific voting information of each voter, anonymity is thereby ensured. As we can obtain the phase information on the single-target qubit via the Wigner function tomography method, the verifiability of the voting result is ensured. Furthermore, Our QVM has scalability in terms of the number of voters and is general in its physical implementation, meaning it is not limited to circuit QED in physical implementation. As long as the multi-control single-target quantum phase gate in our scheme can be realized in other physical systems, the QVM can be implemented. In addition, our scheme involves a simple operation process, requiring only a single-step operation and the reading of a single target qubit. Finally, numerical calculations indicate that even under conservative parameter conditions, our proposed QVM scheme demonstrates a certain feasibility with the current quantum technology.
\begin{acknowledgments}
This work was partly supported by the National Natural Science Foundation of China (NSFC) (12074179, U21A20436), Innovation Program for Quantum Science and Technology (2021ZD0301702), the Natural Science Foundation of Jiangsu Province, China (Grant No. BE2021015-1, BK20232002) and the Jiangsu Funding Program for Excellent Postdoctoral Talent (20220ZB16), and Natural Science Foundation of Shandong Province(Grant No.ZR2023LZH002).
\end{acknowledgments}


\bibliography{Q_V_M}

\end{document}